**Some remarks on Dirac's hole theory versus quantum field theory**

by

Dan Solomon

Rauland-Borg Corporation
3450 W Oakton
Skokie, IL 60076
USA

Phone – 1-847-324-8337
Email: dan.solomon@rauland.com

October 13, 2003

**Abstract**

Dirac's hole theory and quantum field theory are generally considered to be equivalent to each other. However, it has been recently shown by Coutinho et al that this is not necessarily the case. In this article we will discuss the reason for this lack of equivalence and suggest a possible solution.

PACS Nos: 03.65-w, 11.10-z

## I. Introduction.

In two recent papers Coutinho et al [1][2] compare calculations done using Dirac's hole theory with quantum field theory. Specifically, they calculate the change in the vacuum energy due to a perturbing electric potential that is time independent. Hole theory and field theory are suppose to be equivalent, however it is shown in [1] and [2] that when calculations are done different results are obtained. Coutinho suggests that this lack of equivalence is related to the validity of Feynman's prescription to disregard the Pauli exclusion principle in the intermediate states of perturbation theory. It is the purpose of this article to shed further light on this problem.

In order to simplify the analysis and avoid unnecessary mathematical detail Coutinho [2] considers the Dirac equation in 1-1D space-time where the wave function is confined to a region of length 2a. Here we will take the space dimension along the y-axis so that wave function is confined the region $|y| < a$. The Dirac Hamiltonian for a single particle can be written as,

$$H = H_0 + V \tag{1}$$

where $H_0$ is the Dirac Hamiltonian in the absence of interactions and V is an external time independent perturbation. For the 1-1D case,

$$H_0 = -i\sigma_y \frac{d}{dy} + \sigma_x m \text{ and } V = -\sigma_y A_y + A_0 \tag{2}$$

where $(A_0, A_y)$ is a time independent electric potential and $\sigma_x$ and $\sigma_y$ are the usual Pauli matrices. The unperturbed eigenstates satisfy,

$$H_0 \varphi_n^{(0)} = \varepsilon_n^{(0)} \varphi_n^{(0)} \tag{3}$$

where in 1-1D space-time the eigenfunctions $\varphi_n^{(0)}$ are 2-spinors. The $\varphi_n^{(0)}$ are assumed to form an orthonormal basis and satisfy the following relationship (see page 202 of Greiner et al [3]),

$$\sum_{m=-\infty}^{+\infty} \varphi_{m,\alpha}^{(0)}(y)\varphi_{m,\beta}^{(0)}(y') = \delta_{\alpha\beta}\delta(y-y') \tag{4}$$

where $\alpha$ and $\beta$ are spinor indices. If the electric potential is time independent then the wave functions for the perturbed system satisfy,

$$(H_0 + V)\varphi_n = \varepsilon_n \varphi_n \tag{5}$$

In the above expressions the index "n" $(= \pm 1, \pm 2, ...)$ specifies the eigenstates. The energy eigenstates are labeled $0 < \varepsilon_1^{(0)} < \varepsilon_2^{(0)} < ...,$ and $0 > \varepsilon_{-1}^{(0)} > \varepsilon_{-2}^{(0)} > ...,$. All the energy levels are assumed to be discrete and non-degenerate. There is one-to-one correspondence between the unperturbed and perturbed wave functions so that $\varepsilon_n \to \varepsilon_n^{(0)}$ as $V \to 0$.

In Dirac hole theory the vacuum state is the state where each negative energy state is occupied by a single electron and each positive energy state is unoccupied. The energy of the vacuum state is given by summing over the energies of all the negative energy states. The change in energy of a given state 'n' due to the perturbing potential is,

$$\Delta\varepsilon_n = \varepsilon_n - \varepsilon_n^{(0)} \tag{6}$$

Therefore the total change in the vacuum energy is,

$$\Delta E_{hv} = \sum_{n=1}^{\infty} \Delta\varepsilon_{-n} \tag{7}$$

The subscript "hv" means that this is the change in the vacuum energy using hole theory. We shall calculate $\Delta E_{hv}$ to the second order in V using time independent perturbation theory. First define,

$$V_{m,n} = \int_{-a}^{+a} \varphi_m^{(0)\dagger} V \varphi_n^{(0)} dy \tag{8}$$

By the methods of standard time independent perturbation theory (see ref [4] and [5]) the change in the energy of a state 'n' is given by,

$$\Delta\varepsilon_n = \Delta\varepsilon_n^{(1)} + \Delta\varepsilon_n^{(2)} + O\left(V^3\right) \tag{9}$$

where $O\left(V^3\right)$ means terms to the third order in V or higher. The first order term $\Delta\varepsilon_n^{(1)}$ is,

$$\Delta\varepsilon_n^{(1)} = V_{n,n} \tag{10}$$

and the second order term $\Delta\varepsilon_n^{(2)}$ is,

$$\Delta\varepsilon_n^{(2)} = \sum_{\substack{m=-\infty \\ m\neq n}}^{\infty} \frac{\left|V_{m,n}\right|^2}{\left(\varepsilon_n^{(0)} - \varepsilon_m^{(0)}\right)} \tag{11}$$

Use the above relationships in equation (7) to show that the change in the vacuum energy is given by,

$$\Delta E_{hv} = \Delta E_{hv}^{(1)} + \Delta E_{hv}^{(2)} + O\left(V^3\right) \tag{12}$$

where the first order term is,

$$\Delta E_{hv}^{(1)} = \sum_{n=1}^{\infty} V_{-n,-n} \tag{13}$$

and the second order term is,

$$\Delta E_{hv}^{(2)} = \sum_{n=1}^{\infty}\left(\sum_{\substack{m=-\infty \\ m\neq -n}}^{\infty} \frac{\left|V_{m,-n}\right|^2}{\left(\varepsilon_{-n}^{(0)} - \varepsilon_{m}^{(0)}\right)}\right) \tag{14}$$

We can rewrite $\Delta E_{hv}^{(2)}$ as follows,

$$\Delta E_{hv}^{(2)} = Y + X \tag{15}$$

where,

$$Y = -\sum_{n=1}^{\infty}\sum_{m=1}^{\infty} \frac{\left|V_{m,-n}\right|^2}{\left(\varepsilon_{m}^{(0)} - \varepsilon_{-n}^{(0)}\right)} \tag{16}$$

and,

$$X = \sum_{n=1}^{\infty}\left(\sum_{\substack{m=1 \\ m\neq n}}^{\infty} \frac{\left|V_{-m,-n}\right|^2}{\left(\varepsilon_{-n}^{(0)} - \varepsilon_{-m}^{(0)}\right)}\right) \tag{17}$$

## II. Quantum field theory.

In the previous section we derived an expression for the change in the vacuum energy for Dirac's hole theory. In this case each negative energy state is occupied by a single electron. The perturbation changes the energy of each particle. The change in the vacuum energy is just the sum of the change in the energy of each negative energy electron. Now we want to work the same problem using quantum field theory. We shall work in the Schrödinger picture. In this case the field operators are time independent and all changes in the system are reflected in the changes of the state vectors. The field operators are defined by,

$$\hat{\psi} = \sum_{n} \hat{a}_n \varphi_n^{(0)}; \quad \hat{\psi}^{\dagger} = \sum_{n} \hat{a}_n^{\dagger} \varphi_n^{(0)\dagger} \tag{18}$$

where the $\hat{a}_n$ ($\hat{a}_n^\dagger$) are the destruction(creation) operators for a particle in the state $\varphi_n^{(0)}$. The $\varphi_n^{(0)}$ are the same wave functions as in the previous section and satisfy equation (3) with eigenvalues $\varepsilon_n^{(0)}$ which are ordered as described in the previous section. The operators $\hat{a}_n$ and $\hat{a}_n^\dagger$ satisfy the anticommutator relation

$$\hat{a}_m\hat{a}_n^\dagger + \hat{a}_n^\dagger\hat{a}_m = \delta_{mn};\ \text{all other anticommutators=0} \tag{19}$$

The Hamiltonian operator is,

$$\hat{H} = \hat{H}_0 + \hat{V} \tag{20}$$

where,

$$\hat{H}_0 = \int \hat{\psi}^\dagger H_0 \hat{\psi} dy - \xi_{ren} \ \text{and}\ \hat{V} = \int \hat{\psi}^\dagger V \hat{\psi} dy \tag{21}$$

$\xi_{ren}$ is a renormalization constant defined so that the energy of the vacuum state $|0\rangle$ is equal to zero. Note that all integrations are over the range $a \geq y \geq -a$. Using (2) we can write $\hat{V}$ as,

$$\hat{V} = \int \left(-\hat{J}_y A_y + \hat{\rho} A_0\right) dy \tag{22}$$

where $\hat{J}_y$ is current operator and $\hat{\rho}$ is the charge operator. They are defined by,

$$\hat{J}_y = \int \hat{\psi}^\dagger \sigma_y \hat{\psi} dy \ \text{and}\ \hat{\rho} = \int \hat{\psi}^\dagger \hat{\psi} dy \tag{23}$$

Following Greiner (see Chapt. 9 of [3]) define the state vector $|0,\text{bare}\rangle$ which is the state vector that is empty of all particles, i.e.,

$$\hat{a}_n |0,\text{bare}\rangle = 0 \ \text{for all n} \tag{24}$$

The vacuum state vector $|0\rangle$ is defined as the state vector in which all negative energy states are occupied by a single particle. Therefore

$$\left|0\right\rangle = \prod_{n=1}^{\infty} \hat{a}_{-n}^{\dagger} \left|0, \text{bare}\right\rangle \tag{25}$$

From this expression and equations (24) and (19) we have the following relationship,

$$\left\langle 0\right| \hat{a}_m^{\dagger} \hat{a}_n \left|0\right\rangle = \left\langle \begin{array}{l} 0 \text{ if } n > 0 \\ \delta_{mn} \text{ if } n<0 \end{array} \right. \tag{26}$$

The vacuum state satisfies the equation,

$$\hat{H}_0 \left|0\right\rangle = 0 \tag{27}$$

Therefore $\left|0\right\rangle$ is an eigenstate of the operator $\hat{H}_0$ with an eigenvalue $\varepsilon\left(\left|0\right\rangle\right) = 0$.

Additional eigenstates $\left|k\right\rangle$ are produced by acting on $\left|0\right\rangle$ with one or more of the creation operators $\hat{a}_j^{\dagger}$ or destruction operators $\hat{a}_{-j}$ where, in either case, $j > 0$. The effect of the operator $\hat{a}_j^{\dagger}$ ($j > 0$) is to increase the energy of the state by $\varepsilon_j^{(0)}$. The effect of the operator $\hat{a}_{-j}$ ($j > 0$) is to destroy a negative energy electron of energy $\varepsilon_{-j}^{(0)}$. Since this quantity is negative this results in a positive increase in energy. Therefore the eigenstates $\left|k\right\rangle$ satisfy,

$$\hat{H}_0 \left|k\right\rangle = \varepsilon\left(\left|k\right\rangle\right)\left|k\right\rangle \text{ where } \varepsilon\left(\left|k\right\rangle\right) > \varepsilon\left(\left|0\right\rangle\right) = 0 \text{ if } \left|k\right\rangle \neq \left|0\right\rangle \tag{28}$$

Now consider the case where there is a time independent perturbation $\hat{V}$. Using time independent perturbation theory we find that the change in energy of the vacuum state $\left|0\right\rangle$ is,

$$\Delta E_{qv} = \Delta E_{qv}^{(1)} + \Delta E_{qv}^{(2)} + O\left(V^3\right) \tag{29}$$

where,

$$\Delta E_{qv}^{(1)} = \langle 0|\hat{V}|0\rangle \tag{30}$$

and

$$\Delta E_{qv}^{(2)} = \sum_{|k\rangle \neq |0\rangle} \frac{\left|\langle 0|\hat{V}|k\rangle\right|^2}{\left(\varepsilon(|0\rangle) - \varepsilon(|k\rangle)\right)} \tag{31}$$

Note that the subscript 'qv' means that this is the vacuum energy evaluated using quantum field theory as opposed to hole theory.

The only states $|k\rangle \neq |0\rangle$ for which the quantity $\langle 0|\hat{V}|k\rangle$ is nonzero are of the form $|k\rangle = \hat{a}_m^\dagger \hat{a}_{-n}|0\rangle$ where 'n' and 'm' are integers greater than zero. The energy of the state $|k\rangle = \hat{a}_m^\dagger \hat{a}_{-n}|0\rangle$ is $\varepsilon(|k\rangle) = \left(\varepsilon_m^{(0)} - \varepsilon_{-n}^{(0)}\right)$. Use this, along with the fact that $\varepsilon(|0\rangle) = 0$, in (31) to obtain,

$$\Delta E_{qv}^{(2)} = -\sum_{n=1}^{\infty}\sum_{m=1}^{\infty} \frac{\left|\langle 0|\hat{V}\hat{a}_m^\dagger a_{-n}|0\rangle\right|^2}{\left(\varepsilon_m^{(0)} - \varepsilon_{-n}^{(0)}\right)} \tag{32}$$

Next use (21) and (18) to obtain,

$$\langle 0|\hat{V}\hat{a}_m^\dagger \hat{a}_{-n}|0\rangle = \langle 0|\left(\int \hat{\psi}^\dagger V \hat{\psi} dy\right)\hat{a}_m^\dagger \hat{a}_{-n}|0\rangle = \sum_{sr}\langle 0|\hat{a}_s^\dagger \hat{a}_r \hat{a}_m^\dagger \hat{a}_{-n}|0\rangle \int \varphi_s^{(0)\dagger} V \varphi_r^{(0)} dy \tag{33}$$

Use (8) in the above to obtain,

$$\langle 0|\hat{V}\hat{a}_m^\dagger \hat{a}_{-n}|0\rangle = \sum_{sr}\langle 0|\hat{a}_s^\dagger \hat{a}_r \hat{a}_m^\dagger \hat{a}_{-n}|0\rangle V_{s,r} \tag{34}$$

Use (25) and (19) to obtain,

$$\langle 0|\hat{a}_s^\dagger \hat{a}_r \hat{a}_m^\dagger \hat{a}_{-n}|0\rangle \underset{\substack{m>0\\n>0}}{=} \delta_{-n,s}\delta_{mr} \tag{35}$$

Use this result in (34) to yield,

$$\left\langle 0\right|\hat{V}\hat{a}_m^{\dagger}\hat{a}_{-n}\left|0\right\rangle \underset{\substack{m>0\\n>0}}{=} V_{-n,m} \tag{36}$$

Use this in (32) to obtain,

$$\Delta E_{qv}^{(2)} = -\sum_{n=1}^{\infty}\sum_{m=1}^{\infty}\frac{\left|V_{m,-n}\right|^2}{\left(\varepsilon_m^{(0)} - \varepsilon_{-n}^{(0)}\right)} \tag{37}$$

where we have used the fact that $\left|V_{m,-n}\right| = \left|V_{-n,m}\right|$. Recall that $\varepsilon_m^{(0)} > 0$ and $\varepsilon_{-n}^{(0)} < 0$ for m and n greater than zero. Therefore, each term in the sum is positive so that $\Delta E_{qv}^{(2)} < 0$ since, in general, $V_{m,-n}$ is nonzero.

Now compare this result to equation (15). We see that $\Delta E_{qv}^{(2)}$ and $\Delta E_{hv}^{(2)}$ are equivalent only if X is zero. If we assume that the dummy indices 'm' and 'n' in equation (17) can be switched then we can show formally that X is zero. Switching these indices and using the fact that $\left|V_{-m,-n}\right| = \left|V_{-n,-m}\right|$ we obtain $X = -X$. So that it seems reasonable to suggest that $X = 0$ which would make the second order solutions equivalent. However when we work an exact problem in the next section then we will see that this is not the case.

## **III. An Exact Solution.**

In order to further investigate the questions raised by Coutinho we will consider a problem for which we can find an exact solution. When this is done the results of perturbation theory can be compared to the exact results. An example of this has already been discussed by Cavalcanti[6]. Here we shall consider a different example from that presented in [6]. Let the perturbing electric potential be given as,

$$A_0 = 0 \text{ and } A_y = -\frac{\partial \chi(y)}{\partial y} \tag{38}$$

where $\chi(y)$ is a time independent function that satisfies the boundary condition,

$$\chi(a) = \chi(-a) = 0 \tag{39}$$

Other then this $\chi(y)$ is arbitrary. Using this in (5) it is easy to show that,

$$\varphi_n = e^{-i\chi}\varphi_n^{(0)} \text{ and } \varepsilon_n = \varepsilon_n^{(0)} \tag{40}$$

Apply this result to Dirac's hole theory. According to (40) the change in the energy of each state 'n' is zero. Therefore the total change in the vacuum energy must be zero. Now compare this to the result obtained using perturbation theory. For the first order change in the energy for the state 'n' we have, using (10), (2), and (38),

$$\Delta\varepsilon_n^{(1)} = V_{nn} = \int_{-a}^{+a} \varphi_n^{(0)\dagger}\left(\sigma_y \frac{d\chi}{dy}\right)\varphi_n^{(0)} dy \tag{41}$$

Integrate by parts, and use (39), to obtain,

$$\Delta\varepsilon_n^{(1)} = -\int_{-a}^{+a} \chi \frac{d}{dy}\left(\varphi_n^{(0)\dagger}\sigma_y\varphi_n^{(0)}\right)dy \tag{42}$$

Using (2) and (3) it can be shown that,

$$\begin{aligned}\frac{d}{dy}\left(\varphi_n^{(0)\dagger}\sigma_y\varphi_m^{(0)}\right) &= \left(\sigma_y \frac{d\varphi_n^{(0)}}{dy}\right)^{\dagger}\varphi_m^{(0)} + \varphi_n^{(0)\dagger}\left(\sigma_y \frac{d\varphi_m^{(0)}}{dy}\right) \\ &= \left(iH_0\varphi_n^{(0)}\right)^{\dagger}\varphi_m^{(0)} + \varphi_n^{(0)\dagger}\left(iH_0\varphi_m^{(0)}\right) = -i\left(\varepsilon_n^{(0)} - \varepsilon_m^{(0)}\right)\varphi_n^{(0)\dagger}\varphi_m^{(0)}\end{aligned} \tag{43}$$

Use this in (42) to show that $\Delta\varepsilon_n^{(1)} = 0$. Next determine $\Delta\varepsilon_n^{(2)}$. Use the above results to obtain,

$$V_{nm} = \int_{-a}^{+a} \varphi_n^{(0)\dagger}\left(\sigma_y \frac{d\chi}{dy}\right)\varphi_m^{(0)} dy = i\left(\varepsilon_n^{(0)} - \varepsilon_m^{(0)}\right)\int_{-a}^{+a} \varphi_n^{(0)\dagger}\chi\varphi_m^{(0)} dy \tag{44}$$

Use this in (11) to obtain,

$$\Delta\varepsilon_n^{(2)} = -i\sum_{m=-\infty}^{+\infty}\left(\int_{-a}^{+a} \varphi_n^{(0)\dagger}\sigma_y \frac{d\chi}{dy}\varphi_m^{(0)} dy\right)\left(\int_{-a}^{+a} \varphi_m^{(0)\dagger}\chi\varphi_n^{(0)} dy\right) \tag{45}$$

Use the (4) in the above to obtain,

$$\Delta\varepsilon_n^{(2)} = -\frac{i}{2}\int_{-a}^{+a} \varphi_n^{(0)\dagger}\sigma_y \frac{d\left(\chi^2\right)}{dy}\varphi_n^{(0)} dy = \frac{i}{2}\int_{-a}^{+a} \chi^2 \frac{d\left(\varphi_n^{(0)\dagger}\sigma_y\varphi_n^{(0)}\right)}{dy} dy \tag{46}$$

where we have integrated by parts and used the boundary conditions (39). Next use (43) in the above to obtain $\Delta\varepsilon_n^{(2)} = 0$. Therefore the first and second order energy shifts are zero for a given state 'n'. This is in agreement with the exact solution. When these results are used in equation (7) we see that the change in the vacuum energy is zero.

At this point we can come to a conclusion about the term X in (17). If this term is zero then $\Delta E_{hv}^{(2)} = \Delta E_{qv}^{(2)}$ (compare equation (37) with (15)). But as has already been discussed $\Delta E_{qv}^{(2)}$ is nonzero. Therefore X must be nonzero in order that $\Delta E_{hv}^{(2)} = 0$. Now, as discussed above, X can be shown to be zero if we assume that the dummy indices 'n' and 'm' in equation (17) are equivalent. They appear to be equivalent because they are both summations from one to infinity. To understand the problem with this conclusion refer back to equation (14). The outer sum over the index 'n' is the sum over all negative energy states. For a given 'n' the inner sum over 'm' includes both positive and negative terms. The positive terms are those for which $\varepsilon_{-n} > \varepsilon_m$ and the negative terms are those for which $\varepsilon_{-n} < \varepsilon_m$. Due to the fact that there are both negative and positive terms in the sum over ‘m’ it is possible for this sum to be zero which is the case for the example considered above. Next divide up the summation into two parts per equation (15). The first term to the right of the equals sign, Y, is obviously negative and the second term is X. Therefore for the result to be zero X must be positive. When we examine X in equation (17) we see that there are both positive and negative energy terms in the sum. However if we switch dummy indices to obtain $X = -X = 0$ these terms seem to cancel out. The problem with this result is that it depends on the assumption that both summations from 1 to infinity are equivalent. This is only true if we can say that $\infty = \infty$. This is a meaningless statement. To correct this problem we will rewrite (7) as,

$$\Delta E_{Lhv} \underset{L\to\infty}{=} \sum_{n=1}^{L} \Delta\varepsilon_{-n} \tag{47}$$

Instead of summing over all negative energy states from 1 to infinity the sum over the negative energy states is taken from 1 to L where L is considered to be a finite integer that approaches infinity. Note that in the above equation we write $\Delta E_{Lhv}$ instead of $\Delta E_{hv}$ for the change in the vacuum energy to denote that the former quantity is calculated using a limiting process. Therefore equation (14) becomes,

$$\Delta E_{Lhv}^{(2)} \underset{L\to\infty}{=} \sum_{n=1}^{L} \Delta\varepsilon_{-n}^{(2)} = \sum_{n=1}^{L} \left( \sum_{\substack{m=-\infty \\ m\neq -n}}^{\infty} \frac{\left|V_{m,-n}\right|^2}{\left(\varepsilon_{-n}^{(0)} - \varepsilon_{m}^{(0)}\right)} \right) \tag{48}$$

From the above expression it is obvious that $\Delta E_{Lhv}^{(2)} = 0$ for the example considered above where $\Delta\varepsilon_{-n}^{(2)} = 0$. Next divide this expression into to parts to obtain,

$$. \Delta E_{Lhv}^{(2)} \underset{L\to\infty}{=} Y_L + X_L \tag{49}$$

where,

$$Y_L \underset{L\to\infty}{=} -\sum_{n=1}^{L}\sum_{m=1}^{\infty}\frac{\left|V_{m,-n}\right|^2}{\left(\varepsilon_m^{(0)}-\varepsilon_{-n}^{(0)}\right)} \tag{50}$$

and,

$$X_L \underset{L\to\infty}{=} \sum_{n=1}^{L}\left(\sum_{\substack{m=1\\ m\neq n}}^{\infty}\frac{\left|V_{-m,-n}\right|^2}{\left(\varepsilon_{-n}^{(0)}-\varepsilon_{-m}^{(0)}\right)}\right) \tag{51}$$

In the above expressions 'n' and 'm' are not equivalent dummy indices and switching them does not automatically give $X_L = -X_L$. Rewrite $X_L$ as,

$$X_L = X_{L1} + X_{L2} \tag{52}$$

where,

$$X_{L1} \underset{L\to\infty}{=} \sum_{n=1}^{L}\sum_{\substack{m=1\\ m\neq n}}^{L}\frac{\left|V_{-m,-n}\right|^2}{\left(\varepsilon_{-n}^{(0)}-\varepsilon_{-m}^{(0)}\right)} \tag{53}$$

and

$$X_{L2} \underset{L\to\infty}{=} \sum_{n=1}^{L}\left(\sum_{m=L+1}^{\infty}\frac{\left|V_{-m,-n}\right|^2}{\left(\varepsilon_{-n}^{(0)}-\varepsilon_{-m}^{(0)}\right)}\right) \tag{54}$$

Now, in equation (53), switch dummy indices to obtain $X_{L1} = -X_{L1} = 0$. Use this result to obtain,

$$\Delta E_{Lhv}^{(2)} \underset{L\to\infty}{=} Y_L + X_{L2} \tag{55}$$

Now we see how it is possible for the term $\Delta E_{Lhv}^{(2)}$ to be zero. As has already been discussed $Y_L$ is negative. Now refer to (54) and consider $X_{L2}$. The quantity $\left(\varepsilon_{-n}^{(0)} - \varepsilon_{-m}^{(0)}\right)$ is always positive for the case where $m > n$. Therefore each term in the summation in (54) is positive. Therefore $X_{L2}$ is positive and cancels out $Y_L$.

When this problem is worked out in quantum field theory the second order change is given by (37) which is nonzero. In quantum field theory there is no term corresponding to $X_{L2}$ to cancel out the negative term that appears in (37). So we see that there is difference between hole theory and quantum field theory. The reason for this difference will be revealed in the following discussion.

## IV. Gauge invariance and quantum field theory.

We picked the electrical potential given by (38) because this allowed us to find an exact solution to equation (5) for the perturbed system. The energy states for the perturbed system and the unperturbed system are the same so that in hole theory the change in the vacuum energy is zero When this problem is worked out in quantum field theory it is found that the second order change in the vacuum energy is nonzero. The question that we need to address is why do hole theory and field theory yield different results and which method yields the correct result?

Now the electrical potential given by (38) has a property which will be useful when we consider quantum field theory. In 1-1D space-time the electric field is given in terms of the electric potential by,

$$E_y = -\left(\frac{\partial A_y}{\partial t} + \frac{\partial A_0}{\partial y}\right) \tag{56}$$

When the electric potential of (38) is used in the above it is seen that the electric field is zero because $\chi(y)$, and thereby $A_y$, is time independent and $A_0$ is zero. Therefore the electric field is the same for the perturbed system as for the unperturbed system. This means that the electric potentials are related by a gauge transformation. A gauge transformation is a change in the electric potential that produces no change in the electric field. Dirac field theory is assumed to be gauge invariant [7]. This means a change in the gauge does not produce a change in any physical observables. These include current and charge expectation values as well as the difference in energy between different physical states.

A gauge transformation is a type of symmetry transformation [2]. According to Weinberg [8] a symmetry transformation is a change in our point of view that does not change the results of possible experiments. Let two observers S and $S_g$ be related by a symmetry transformation. For the observer S let the state vector $|\Omega\rangle$ correspond to a given physical system. For $S_g$ we will designate the state vector that corresponds to the same physical system by $|\Omega_g\rangle$. Now how is $|\Omega_g\rangle$ related to $|\Omega\rangle$? According to [8] a symmetry transformation must preserve the inner product between two different state vectors, that is, if $|\Omega\rangle$ and $|\Omega'\rangle$ are two different state vectors in S and $|\Omega_g\rangle$ and $|\Omega'_g\rangle$ are the corresponding state vectors in $S_g$ then $\langle\Omega|\Omega'\rangle = \langle\Omega_g|\Omega'_g\rangle$. This means that $|\Omega\rangle$ and $|\Omega_g\rangle$ are related by a unitary transformation so that $|\Omega_g\rangle = \hat{U}|\Omega\rangle$ where $\hat{U}$ is a unitary operator.

Let observer S operate in a gauge where the electrical potential is zero. Let $S_g$ operate in a gauge where the electric potential is given by (38). Since the electric field is the zero in both cases both observers are describing the same physical system. First consider observer S. Recall equation (28) which we rewrite below for convenience,

$$\hat{H}_0 \left| k \right\rangle = \varepsilon\left(\left| k \right\rangle\right)\left| k \right\rangle \text{ where } \varepsilon\left(\left| k \right\rangle\right) > \varepsilon\left(\left| 0 \right\rangle\right) = 0 \text{ if } \left| k \right\rangle \neq \left| 0 \right\rangle \tag{57}$$

The corresponding equation from the point of view of $S_g$ is,

$$\left(\hat{H}_0 + \hat{V}\right)\left| k_g \right\rangle = \varepsilon\left(\left| k_g \right\rangle\right)\left| k_g \right\rangle \text{ where } \varepsilon\left(\left| k_g \right\rangle\right) > \varepsilon\left(\left| 0_g \right\rangle\right) \text{ if } \left| k_g \right\rangle \neq \left| 0_g \right\rangle \tag{58}$$

where, from (22) and (38),

$$\hat{V} = \int \hat{J}_y \frac{\partial \chi}{\partial y} dy \tag{59}$$

Note that $\left| k \right\rangle$ describes a given physical system from the point of view of S and $\left| k_g \right\rangle$ describes the same physical system from the point of view of $S_g$ where $\left| k_g \right\rangle = \hat{U}\left| k \right\rangle$. Also, even though, in general, $\left| k_g \right\rangle \neq \left| k \right\rangle$, all physical observables described from both points of view must be the same.

Now according to QFT the eigenstates $\left| k \right\rangle$ form an orthonormal basis and any arbitrary state vector $\left| \Omega \right\rangle$ can be expanded in terms of this basis. Given this, and using (57), it is easy to show that,

$$\left\langle \Omega \right| \hat{H}_0 \left| \Omega \right\rangle > \left\langle 0 \right| \hat{H}_0 \left| 0 \right\rangle = 0 \text{ for all } \left| \Omega \right\rangle \neq \left| 0 \right\rangle \tag{60}$$

The corresponding relationship according to $S_g$ is,

$$\left\langle \Omega_g \right| \left(\hat{H}_0 + \hat{V}\right) \left| \Omega_g \right\rangle > \left\langle 0_g \right| \left(\hat{H}_0 + \hat{V}\right) \left| 0_g \right\rangle \text{ for all } \left| \Omega_g \right\rangle \neq \left| 0_g \right\rangle \tag{61}$$

Now due to the fact that $\left| \Omega \right\rangle$ is arbitrary we may consider the case where $\left| \Omega \right\rangle = \left| 0_g \right\rangle$. Substitute this into equation (60) to obtain,

$$\left\langle 0_g \right| \hat{H}_0 \left| 0_g \right\rangle > \left\langle 0 \right| \hat{H}_0 \left| 0 \right\rangle \tag{62}$$

Similarly, in equation (61), let $\left| \Omega_g \right\rangle = \left| 0 \right\rangle$ to obtain,

$$\left\langle 0\right|\left(\hat{H}_0+\hat{V}\right)\left|0\right\rangle > \left\langle 0_g\right|\left(\hat{H}_0+\hat{V}\right)\left|0_g\right\rangle \tag{63}$$

The expectation value of the current in the vacuum state is assumed to be zero. So that the observer S can write,

$$\left\langle 0\right|\hat{J}_y\left|0\right\rangle = 0 \tag{64}$$

However from the principle of gauge invariance we have a corresponding equation for the observer $S_g$ which is,

$$\left\langle 0_g\right|\hat{J}_y\left|0_g\right\rangle = 0 \tag{65}$$

Use this result in (59) to obtain,

$$\left\langle 0\right|\hat{V}\left|0\right\rangle = \left\langle 0_g\right|\hat{V}\left|0_g\right\rangle = 0 \tag{66}$$

Use this in (63) to yield,

$$\left\langle 0\right|\hat{H}_0\left|0\right\rangle > \left\langle 0_g\right|\hat{H}_0\left|0_g\right\rangle \tag{67}$$

Note that there is a contradiction between this result and equation (62). This reflects an inconsistency in quantum field theory between the assumption of gauge invariance and equation (60). This problem was discussed in [9] where it was shown for quantum field theory to be gauge invariant there must exist quantum states whose energy is less than that of the vacuum state.

In [9] it was shown that this could be done by redefining the vacuum state as follows,

$$\left|0_L\right\rangle \underset{L\to\infty}{=} \prod_{n=1}^{L} \hat{a}_{-n}^{\dagger}\left|0,\text{bare}\right\rangle \tag{68}$$

According to this definition the band of vacuum states with energies from $\varepsilon_{-1}$ to $\varepsilon_{-L}$ are occupied by a single electron. All positive energy states are unoccupied. All states with energy less than $\varepsilon_{-L}$ are also unoccupied. This differs from the standard definition of the vacuum state $|0\rangle$ given by equation (25). The difference is due to the fact that the lower edge of the negative energy band is defined using a limiting procedure.

Eigenstates $|k_L\rangle$ are produced by acting on $|0_L\rangle$ with creation operators $\hat{a}_j^\dagger$ where $j > 0$ or $j < -L$ and with destruction operators $\hat{a}_{-j}$ with $L > j \geq 1$ . The effect of operator $\hat{a}_j^\dagger$ where $j < -L$ is to place an electron in one of the unoccupied negative energy states that exist underneath the band of occupied negative energy states. This creates a quantum states with less energy then the vacuum state $|0_L\rangle$. As discussed above and in ref. [9] this is a necessary requirement for quantum field theory to be gauge invariant.

Now refer to equation (31) and substitute $|0_L\rangle$ for $|0\rangle$ to obtain,

$$\Delta E_{Lqv}^{(2)} \underset{L\to\infty}{=} \sum_{|k_L\rangle \neq |0_L\rangle} \frac{\left|\langle 0_L|\hat{V}|k_L\rangle\right|^2}{\left(\varepsilon\left(|0_L\rangle\right) - \varepsilon\left(|k_L\rangle\right)\right)} \tag{69}$$

The states that make a nonzero contribution to the above sum are of the form $|k_L\rangle = \hat{a}_m^\dagger \hat{a}_{-n} |0_L\rangle$ where $L \geq n \geq 1$ and 'm' is either a positive integer or an integer less than $-L$. The energy of the state $|k_L\rangle = \hat{a}_m^\dagger \hat{a}_{-n} |0_L\rangle$ is $\varepsilon\left(|k_L\rangle\right) = \left(\varepsilon_m^{(0)} - \varepsilon_{-n}^{(0)}\right)$. Use this, along with $\varepsilon\left(|0_L\rangle\right) = 0$, in (69) to obtain,

$$\Delta E_{Lqv}^{(2)} \underset{L\to\infty}{=} \sum_{n=1}^{L} \left( -\sum_{m=1}^{\infty} \frac{\left|V_{m,-n}\right|^2}{\left(\varepsilon_m^{(0)} - \varepsilon_{-n}^{(0)}\right)} + \sum_{m=L+1}^{\infty} \frac{\left|V_{-m,-n}\right|^2}{\left(\varepsilon_{-n}^{(0)} - \varepsilon_{-m}^{(0)}\right)} \right) \tag{70}$$

Now refer back to the discussion leading up to equation (55). From this we see that the first term in the above equation corresponds to $Y_L$ and the second term to $X_{L2}$ so that,

$$\Delta E_{Lqv}^{(2)} = \Delta E_{Lhv}^{(2)} \tag{71}$$

Therefore hole theory and field theory yield identical results when the vacuum state is defined per equation (68).

## V, Conclusion.

We have examined the difficulties involved in determining the second order change of the vacuum energy using Dirac's hole theory and quantum field theory. In examining this problem we have used a perturbing electric potential for which we know the solution on both mathematical and theoretical grounds. When the electric potential is given by equation (38) the exact solution to the Dirac equation is easily obtained. In this case the change in the energy states are zero so that the change in the vacuum energy must be zero. We show that we can achieve the same result using perturbation theory, however there is some potential ambiguity if the order of the summations is switched. This ambiguity can be resolved if the sum over the negative energy states is done using the limiting procedure described in Section III.

When we apply perturbation theory to quantum field theory we initially derive equation (37). This result yields the conclusion that the second order change in the vacuum energy must always be negative even for the potential given by (38). Therefore we appear to have a contradiction between the results of hole theory and field theory. This contradiction is shown to be due a problem in field theory in regards to the way that the vacuum state is defined. When the vacuum state is defined according to the standard definition (per equation (25)) then equation (37) is negative because all quantities of form $\varepsilon(|0\rangle) - \varepsilon(|k\rangle)$ are less than zero. It is shown in Section IV that there are theoretical difficulties associated with this definition of the vacuum state. In particular it was shown that for quantum field theory to be gauge invariant there must exist quantum states with less energy than the vacuum state. This result was derived previously in [9]. When the vacuum state is defined according equation (68) then we find agreement between Dirac's hole theory and quantum field theory so that the questions raised by Coutinho et al are resolved.